\documentstyle[aps,prb,graphicx]{revtex}
\newif\ifbuidln
\buidlntrue
\newcommand{\senfschnitt}{\theta_\Lambda}
\newcommand{\renG}{{\tilde{\gamma}}}
\newcommand{\renV}{{\tilde{V}}}
\begin{document}
\title{Flow of the quasiparticle weight in the $N$-patch renormalization group scheme}
\author{Carsten Honerkamp$^{1}$ and Manfred Salmhofer$^{2}$}
\address{$^{1}$ Department of Physics, Massachusetts Institute of Technology, Cambridge
MA 02139, USA \\
$^{2}$ Theoretische Physik, Universit\"at Leipzig, D-04109 Leipzig, Germany} 
\date{December 3, 2002}
\maketitle
\begin{abstract} 
Using the $N$-patch renormalization group method
we investigate the flow of the quasiparticle weight in 
one--dimensional, weakly two--dimensional and fully two--dimensional
Hubbard models. 
In one dimension we reproduce the Luttinger exponent 
that describes the disappearance of the quasiparticle peak towards lower scales. 
Further we analyze effects of the band structure, interchain hopping 
and the flow of the interactions in quasi-one-dimensional models. 
For the two-dimensional case we study how the suppression of 
the quasiparticle weight affects the flow to strong coupling.
We find that the flow to strong coupling remains essentially unchanged. 
This strengthens the evidence for $d$-wave superconductivity in the weakly 
repulsive Hubbard model. 
\end{abstract}
\section{Introduction}
Renormalization group methods have contributed in many ways 
to the understanding of the low energy properties of interacting fermion systems. 
Early work using these methods focused on one-dimensional systems\cite{solyom}; 
mathematical studies have been done for weakly coupled systems 
in one\cite{BGPS} and two dimensions \cite{FMRT,FST,salmcrg,DR,FKT}. 
General properties of interacting electrons were discussed in \cite{shankar,chen}. 
The exact RG schemes were used to prove the existence 
of the Landau-Fermi liquid in two dimensions 
under certain conditions\cite{salmcrg,DR,FKT}. 

In the course of the last few years, the method has been applied to 
highly anisotropic two-dimensional models like the Hubbard model 
on the square lattice and variants thereof%
\cite{zanchi,halboth,hsfr,sh,tsai,honedhd,zanchiZ,tflow,binz} 
without any crude simplifications of the dispersion relation and phase space. 
The approximate RG schemes used in these studies are derived from 
exact renormalization group
equations\cite{salmcrg,sh,wegner,polchinski,wetterich,salmhofer} 
that can be applied to wide variety of models 
and that allow a clearer view on the approximations necessary in practical
calculations. 

In these approximate studies in two dimensions, 
the flow of the self energy has up to now mainly been neglected,
with the exception of the following studies.
For the $t,t'$ Hubbard model the Fermi surface flow was included 
dynamically in the RG flow\cite{hsfr} (see the Appendix of that paper). 
It was found that at a fixed density close to the van Hove density,
$t'$ gets reduced in the flow. 
The quasiparticle scattering rate was estimated in 
the $t,t'$ Hubbard model\cite{honedhd}.
The quasiparticle weight was calculated for the special case 
of the half-filled band by Zanchi\cite{zanchiZ}. He found a 
substantial reduction of the weight at the saddle points 
$(0,\pm\pi)$ and $(\pm\pi,0)$. 

In the following, we discuss the question of selfenergy effects in a wider context,
to bring out the points relevant for this paper. 
The selfenergy shifts the Fermi surface and it changes the Fermi velocity 
and the quasiparticle weight, as well as the curvature of the Fermi surface
and the quasiparticle scattering rate. For weakly coupled systems 
with curved Fermi surfaces and without van Hove singularities, 
detailed phase space arguments were applied to prove that the curvature
of the Fermi surface develops no singularity, the quasiparticle weight
remains close to one and the correction to the Fermi velocity is small
\cite{FST,salmcrg,DR}. In particular, this flow does not change the 
instabilities of the Fermi liquid towards a superconducting state,
which shows up as an initially marginal growth of the four--point function 
that leads to a divergence at a scale related to the critical temperature. 
Moreover, the old question of continuity\cite{Kohn} of the results of perturbation
theory
in the limit $T \to 0$ was solved: it was shown in Ref.\ \onlinecite{FKST}
that the problem of "anomalous diagrams" in Ref.\ \onlinecite{Kohn} arose
only because of an insufficient renormalization procedure (which amounts to
the attempt of describing the shift of the Fermi surface merely by 
a shift in the chemical potential). 

The deformation of the Fermi surface was calculated by Halboth and Metzner
in perturbation theory in the Hubbard model\cite{HM-FSpert}; 
they found that there is a density $\rho \approx 0.6$, 
below which the Fermi surface tends to become rounder 
and above which it tends to become more squarelike. 
We note that this density value can be understood geometrically 
as the smallest one where Umklapp processes first contribute to the
two--loop selfenergy; at this density the singularity analysis
of Ref.\ \onlinecite{FST} changes as well. Recently it was pointed out that 
a Fermi surface close to the van Hove points is in principle unstable 
towards a spontaneous breaking of the fourfold symmetry\cite{HMpom,NMpom}.

A curved Fermi surface away from the critical points of the dispersion relation
is a necessary condition for Fermi liquid behavior.  
In presence of van Hove singularities and in situations with small curvature, 
such as very close to half--filling at $t'=0$ and in quasi--one--dimensional 
systems, the selfenergy can be expected to be relevant. 
%
The prototypical case of a perfectly flat Fermi surface is
(apart from half--filling in two dimensions, where the situation is further
complicated by the presence of van Hove singularities)
the one--dimensional fermion system.  
Simple descriptions of the flow towards the Luttinger liquid within the Wilsonian RG
concept 
have been given for example by Metzner et al. \cite{metzner1D} 
or the recent comprehensive review by Bourbonnais\cite{bourbonnais}. 
A mathematical proof that the flow of the quasiparticle weight leads to the 
Luttinger exponents was given by Benfatto et al.\cite{BGPS}.
Recently Busche et al.\cite{busche} rederived the flow of 
the quasiparticle weight $Z$ in the 1D Luttinger model 
and even obtained the spectral function of the Luttinger fixed point. 
Thus the functional RG schemes give a detailed description of 
Luttinger liquid physics. 
One important advantage of this method is that it neither requires exact scaling laws
nor relies on the peculiarities of one spatial dimension,
and that it applies to a wide class of model Hamiltonians.

From the RG point of view, the  existence of the Luttinger liquid fixed point 
hinges on the cancellations to all orders between the particle--hole terms
at $2k_F$ and the particle--particle terms at $0$. This cancellation is there
already in lowest order, where the wave function renormalization still does not
enter. It is a one--dimensional phenomenon:
in higher dimensions, it is absent, and the interactions almost always flow 
to strong coupling, i.e.\ diverge at a positive energy scale. 
Already before this point, the RG flow in the fermionic variables breaks down.
This signals the possible opening of a gap in the fermionic excitation spectrum,
the simplest case being superconductivity.
A key question is if the selfenergy effects, when fed back  
into the flow equations of the four--point interaction, 
can remove such instabilities altogether. 
In particular, a strong suppression
of the quasiparticle weight might prevent a flow to strong coupling,
and thus instabilities with ordering tendencies, at all. 
For curved Fermi surfaces, this has proved not to be the case \cite{FST,salmcrg,DR}, 
but the situation is less simple for quasi-1D systems\cite{kishine,ferraz} 
and models with van Hove singularities on the Fermi
surface\cite{zanchiZ,dzyaloshinski,irkhin}.

In the present article we address this question using 
the numerical $N$--patch implementation of the RG scheme
for the 1PI functions\cite{sh}.
We incorporate the second order flow of the selfenergy in a way similar to 
the one--dimensional study by Busche et al.\cite{busche}, 
but we also include the flow of the renormalized interactions,
which is relevant in more than one dimension.

The main results are as follows. 
Our calculation correctly reproduces the suppression of the quasiparticle weight 
and the exponent of the Luttinger model in the one--dimensional Hubbard model. 
In quasi--1D systems with curved Fermi surfaces the flow to strong coupling 
remains almost unaffected by renormalization of the quasiparticle weight,
in the parameter regime we study. The same holds for the fully two--dimensional 
system: the suppression of the quasiparticle weight does not prevent 
the flow to strong coupling. This implies that the $d$-wave superconductivity found in the weakly repulsive Hubbard model\cite{zanchi,halboth,hsfr,tsai} close to half filling is not destroyed by the inclusion of these selfenergy effects.  
Another piece of information that is provided by our scheme 
is the anisotropy of the quasiparticle weight around the Fermi surface. 
At temperatures above the flow to strong coupling 
this information can be used for a Fermi liquid description of the normal state. 
Our finding is that in the simple Hubbard model the quasiparticle weight varies only mildly around the Fermi surface and no anomalous effects should be expected based on that. 

  
\section{The method}
The renormalization group equations for the 1-particle irreducible (1PI) vertex
function have been described in several works\cite{hsfr,sh,wetterich,kopietz}. 
The basic input for these equations is to assume that the quadratic part of 
the original action depends continuously on a certain parameter, here called $\Lambda$.
The RG equations then describe the evolution of the vertex functions of the system 
when this parameter is varied. 
In most applications $\Lambda$ is taken to be an infrared energy cutoff 
(or, similarly, momentum shells are integrated out); 
under certain conditions however it may be appropriate to use 
the physical temperature\cite{tflow}. 
Although the latter choice is intuitively appealing and preferable 
when investigating if ferromagnetic instabilities occur\cite{tflow}, 
we will use here the conventional IR cutoff scheme, 
because it is simpler and requires less numerical effort 
in the $N$-patch implementation, and because here we shall only study
parameter regions where ferromagnetism was found to be absent\cite{tflow}. 

\begin{figure}
\begin{center} 
\ifbuidln\includegraphics[width=.35\textwidth]{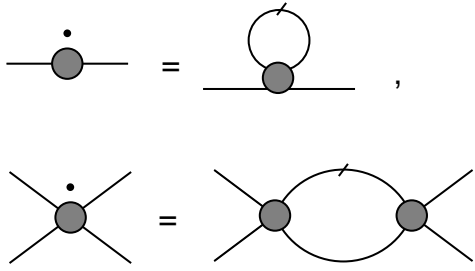}\fi
\end{center} 
\caption{Flow of two-point and four-point vertex.}
\label{oneloopeqs}
\end{figure}

In the IR cutoff RG the quadratic part $Q$ of the action, 
in our case $Q( i \omega, \vec{k}) = i \omega - \epsilon_{\vec{k}}$, 
is supplemented with a cutoff function $\chi_\Lambda (\vec{k})$,
\[ 
T \sum_{i \omega} \sum_{\vec{k}} 
\bar{\psi} ( i \omega, \vec{k}) Q( i \omega, \vec{k}) \psi ( i \omega, \vec{k})  
\longrightarrow \, 
T  \sum_{i \omega} 
\sum_{\vec{k}} \bar{\psi} ( i \omega, \vec{k}) Q( i \omega, \vec{k}) \psi ( i \omega,
\vec{k}) \,  
\chi_\Lambda^{-1} (\vec{k}) \, .  
\] 
We take 
$\chi_\Lambda(\vec{k}) = K (\epsilon_{\vec{k}} - \Lambda)+ K (-\epsilon_{\vec{k}} -
\Lambda)$ 
where $K(x)$ is a fixed smooth function that increases from $0$ to $1$ 
in a small interval of length $2\eta $ around $x=0$. 
Later we shall take the limit $\eta \to 0$, which gives a step function cutoff. 
The function $K$ and the stepfunction limit are discussed in the Appendix.
Thus $\chi_\Lambda(\vec{k})^{-1}$ is unity for band energy $|\epsilon_{\vec{k}}| >
\Lambda$ 
but gets very large for  $|\epsilon_{\vec{k}} | < \Lambda$, thus strongly suppressing 
low energies and effectively restricting the integration 
over the fermionic modes to those above $\Lambda$.  
Taking the derivative with respect to $\Lambda$ one obtains a functional RG equation 
for the generating functional of the connected correlation functions and its 
Legendre transformation, the effective action. 
An expansion in the fields yields an hierarchy of 
coupled differential equations for the 1PI $m$--point vertex functions.
The hierarchy is infinite because the derivative of the $m$-point vertex depends on  
the $(m+2)$--point vertex. 
We truncate the hierarchy by setting the 1PI 6--point function equal to zero,
and are left with the two equations presented graphically in Fig. \ref{oneloopeqs}. 
The first equation determines the flow of the self energy $\Sigma_\Lambda (i \omega,
\vec{k})$
and the second that of the two--particle interaction, i.e.\ the four-point vertex.
The barred internal line in Fig. \ref{oneloopeqs} denotes a single scale propagator
\begin{equation}\label{singlescale}
S_\Lambda (i \omega, \vec{k}) = 
- G_\Lambda (i \omega, \vec{k}) 
\frac{\partial}{\partial\Lambda} Q_\Lambda (i \omega, \vec{k}) G_\Lambda (i \omega,
\vec{k})  
= \frac{(i \omega - \epsilon_{\vec{k}} )}%
{(i \omega - \epsilon_{\vec{k}} - \chi_\Lambda (\vec{k})
\Sigma_\Lambda (i \omega, \vec{k}) )^2} 
\frac{\partial}{\partial\Lambda}{\chi_\Lambda (\vec{k})} \, ,
\end{equation}
while the internal line without bar stands for the interacting 
one-particle Green function at the given scale,
\begin{equation}  \label{G2}
G_\Lambda (i \omega, \vec{k}) 
=  \frac{\chi_\Lambda (\vec{k})}{i \omega - \epsilon_{\vec{k}} 
- \chi_\Lambda (\vec{k})\Sigma_\Lambda (i \omega, \vec{k}) }  
\, .
\end{equation} 

In this paper we consider the limit of a sharp cutoff, $K(x) = \theta (x)$. 
We discuss in the Appendix how this limit is taken; the result is
\begin{equation}\label{Galapeno} 
G_\Lambda (i \omega, \vec{k}) 
=  \frac{\senfschnitt  (\vec{k})}{i \omega - \epsilon_{\vec{k}} 
- \Sigma_\Lambda (i \omega, \vec{k}) }  
\end{equation} 
and 
\begin{equation}\label{Salapeno}
S_\Lambda (i \omega, \vec{k}) = 
- \frac{\delta(\epsilon_{\vec{k}}-\Lambda) + \delta(\epsilon_{\vec{k}}+\Lambda)}%
{i \omega - \epsilon_{\vec{k}} - \Sigma_\Lambda (i \omega, \vec{k})}
=
\frac{1}{i \omega - \epsilon_{\vec{k}} - \Sigma_\Lambda (i \omega, \vec{k}) } \; 
\frac{\partial}{\partial\Lambda} \senfschnitt  (\vec{k})
\end{equation}
with $\senfschnitt (\vec{k}) = \theta (\epsilon_{\vec{k}} - \Lambda) 
+\theta (-\epsilon_{\vec{k}} - \Lambda)$.
When inserting these expressions in the flow equations, undefined quantities like 
$\theta(E-\Lambda )\delta(E-\Lambda)$ can appear at special values of the shift
momentum 
$q$ (e.g.\ at $q=0$). The sharp cutoff limit of all such terms is,
however, well--defined and unique. We discuss in some detail in the appendix 
the limiting rule for avoiding these (seeming) ambiguities.

By spin rotation symmetry, the coupling function is 
\[ 
\gamma^{(4)}_{s_1s_2s_3s_4} ( {k}_1, {k}_2, {k}_3, {k}_4 ) = 
\delta_{s_1,s_3} \delta_{s_2,s_4} V_\Lambda ({k}_1, {k}_2, {k}_3) - 
\delta_{s_2,s_3} \delta_{s_1,s_4} V_\Lambda ({k}_2, {k}_1, {k}_3)  
\]
(with $k_4$ fixed by momentum conservation in terms of $k_1,k_2$ and $k_3$).
Here $k_i = (\omega_i,\vec{k}_i)$ includes frequency and spatial momentum. 

In terms of an expansion in the bare coupling function, the above--mentioned truncation
is
exact up to second order. However, the RG provides more than just a second--order
calculation: the evolution of the interaction and the selfenergy is continually 
fed back into the RG differential equation. This effectively sums up contributions from
arbitrarily high order and thus produces a scale--dependent resummation of 
perturbation theory. We note that the above--mentioned truncation does not correspond
to an
expansion to a fixed loop order: the flow equations appear to be one--loop, 
but they also take into account two--loop effects by iteration. Since we are not doing
a 
semiclassical expansion in orders of $\hbar$, this is not a problem by itself.
The relevant question is when higher orders significantly change the flow.
They certainly do so when the coupling functions get too large; a detailed 
discussion of what "large" means is given in Ref.\ \onlinecite{sh}.

By power-counting arguments, the leading part of the flow is given by 
the interaction processes between particles close to the Fermi surface,
that is, with frequency close to zero and spatial momenta on the Fermi surface. 
Thus we drop the frequency dependence of the four-point vertex, 
so that $V_\Lambda ({k}_1, {k}_2, {k}_3) = V_\Lambda (\vec{k}_1, \vec{k}_2,
\vec{k}_3)$,
and calculate $V_\Lambda (\vec{k}_1, \vec{k}_2, \vec{k}_3)$ 
with wavevectors  $\vec{k}_1, \vec{k}_2$ and $ \vec{k}_3$ on the  Fermi surface.
We discretize the remaining angular dependence by introducing patches around $N$
wavevectors 
on the Fermi surface and approximate the coupling function by its $O(N^3)$ values
at the patch wavevectors. 
This device of angular patches (or "sectors") has been used in various works.
It is the basic method for mathematical studies of 2D fermion systems,
where $N$ is chosen as $\Lambda^{-1/2}$ at scale $\Lambda$, 
i.e.\ it increases at low energies, as one approaches the Fermi surface. 
The strong constraints implied by momentum and energy conservation in 
two dimensions then yield an intrinsic $1/N$ expansion\cite{FMRT,shankar,chen}.
As a concrete calculational tool this $N$-patch technique was first used 
by Zanchi and Schulz\cite{zanchi} and then in a number of subsequent
works\cite{halboth,hsfr,tsai,honedhd,zanchiZ,tflow,binz}.
The projection to the Fermi surface suffices to obtain the leading coupling function
flow
but it would be incorrect to assume that the full 1PI four--point function
is well approximated by the projection to the Fermi surface. 
To obtain observables, one has to study the flow of response
functions\cite{zanchi,halboth,hsfr},
or, more generally, the flow of functions with general external momenta.

\begin{figure}
\begin{center} 
\ifbuidln\includegraphics[width=.15\textwidth]{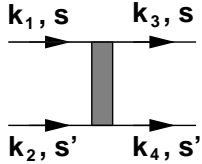}\fi
\end{center} 
\caption{The coupling function $V_\Lambda(\vec{k}_1, \vec{k}_2, \vec{k}_3)$. The spin
indices belonging to wavevectors $\vec{k}_1$ and $\vec{k}_3$ (and  $\vec{k}_2$ and
$\vec{k}_4$) are the same.}
\label{couplingfun}
\end{figure} 

The quasiparticle residue that we want to calculate is determined by the frequency 
dependence of the selfenergy. Because we do not use a cutoff in the frequencies, 
we would in principle have to calculate the self energy over the whole frequency axis. 
However, the singularity of the propagator is at zero frequency, 
so the behavior of the selfenergy near to zero frequency is the most important. 
We calculate the frequency derivative at zero imaginary frequency 
and at the Fermi surface as
\begin{equation} \label{freqdiff}
\partial_\omega \Sigma_\Lambda (0, \vec{k}_F)  
= \frac{\Sigma_\Lambda (i \pi T, \vec{k}_F) - \Sigma_\Lambda (-i \pi T, \vec{k}_F)}{2
\pi T} 
\, , 
\end{equation}
where the denominator is the difference between the two smallest Matsubara frequencies 
$\omega_{\pm 1} = \pm \pi T$
(here $\vec{k}_F$ denotes the projection of $\vec{k}$ to the Fermi surface).
We then determine the quasiparticle weight as  
\begin{equation} \label{Zdef}
Z_\Lambda (\vec{k}) 
=  \left[ 1 + i \partial_\omega \Sigma_\Lambda (0, \vec{k}_F) \right]^{-1}  .
\end{equation} 
and write 
\begin{eqnarray}\label{Zpropagators}
G_\Lambda (i \omega, \vec{k} ) &=& 
\frac{Z_\Lambda ( \vec{k} )}%
{i \omega - \epsilon_{\vec{k}}} \; \senfschnitt (\vec{k})
=
Z_\Lambda ( \vec{k} ) \; C_\Lambda (\vec{k})
\, ,  
\\ 
S_\Lambda (i \omega, \vec{k} ) 
&=& 
\frac{Z_\Lambda ( \vec{k} )}{i \omega - \epsilon_{\vec{k}}} \,
\frac{\partial}{\partial\Lambda} \senfschnitt (\vec{k}) 
=
Z_\Lambda ( \vec{k} ) \; \frac{\partial}{\partial\Lambda} C_\Lambda (\vec{k})
\,  . 
\end{eqnarray}
The above involves a number of approximations. 
As already discussed, we have approximated the quasiparticle weight 
for a particle at band energy $\epsilon_{\vec{k}}$ 
by the quasiparticle weight on the Fermi surface at the RG scale 
$\Lambda $. 
This effectively reduces the total weight to $Z_\Lambda$ because high frequency
modes also get this weight. 
Since we are interested in the low energy properties, this plays no important role. 
We have also neglected the Fermi surface shift $\Sigma_\Lambda (0,\vec{k}_F)$.
Moreover, in writing (\ref{Zpropagators}), 
we have assumed that the correction to the $\vec{k}$--dependence of the 
dispersion relation (involving $\nabla \Sigma_\Lambda (0,\vec{k}_F)$) 
is of the same order of magnitude as $Z_\Lambda$ so that we can effectively 
write $ \epsilon_{\vec{k}}$ in the denominator once we have put 
$Z_\Lambda$ in the numerator of the propagators. 
This is correct in the one--dimensional case because there, the dependence on $k_0$ and
on $k$ is similar; 
in 2D it is a further assumption. An investigation of this assumption 
is underway, as well as a more detailed calculation of the Fermi surface shift
extending that of Ref.\ \onlinecite{hsfr}. 
The quasiparticle scattering rate in the 2D case was considered in Ref.
\onlinecite{honedhd} 
and turned out to be reasonably small until the couplings exceed the perturbative
range.

As indicated above, the frequency dependence of the four--point vertex can be dropped 
in determining the leading part of the flow, but not from the full four--point
function.
In particular, it is obvious from the form of the equation for the selfenergy 
that the external momentum does not enter the internal line,
hence the frequency dependence of the selfenergy must come from a frequency dependence
of the four--point function. Again we employ an approximation that is exact to second
order --
like in Ref. \onlinecite{honedhd} we reinsert the integrated form of the equation for
the four--point function in the selfenergy equation. This gives a two--loop diagram,
in which we now again approximate the two appearing full four--point functions
$\gamma^{(4)}$ 
by coupling functions $V_{\Lambda'}$. The right 
hand side of the differential equation for the selfenergy is now nonlocal in $\Lambda$:
the change of four-point vertex and selfenergy at scale $\Lambda$ 
involves four-point vertices and selfenergies at scales $\Lambda' \ge \Lambda$. In the next section we will compare this nonlocal equation with its local approximation where four-point vertex and self-energy at scale $\Lambda'$ are approximated with the ones at $\Lambda$.   

This iteration of the solution of the RGDE is a general method 
that can also be applied to the full system of equations: 
setting the 1PI $2m$--point function equal to zero means restricting the 
$2m$--point function to tree diagrams in the bare interaction. If the bare
interaction is a four--point interaction, these trees are of order 
$r=m-1$ in the four--point coupling. The resulting truncated system of equations
can be solved by iterative substitution as above.
Thus the truncation of setting the 1PI $2m$--point function equal to zero 
is exact to order $m-1$ in the coupling, and indeed we will reproduce the 
Luttinger liquid exponents to second order in our calculation. 
By the same method one can get higher order corrections to the Luttinger exponents
if one truncates at a higher $m$. 
 
We note that this method to reconstruct higher-loop contributions to the selfenergy is less approximate than the scheme used in Ref. \onlinecite{zanchiZ}. The latter assumes that the forward scattering occurring in the one-loop diagrams for the selfenergy can be decomposed in parts that only depend on  the total momentum or the momentum transfer (which may be less accurate for anisotropic systems) and interchanges scale and frequency dependences of the scattering vertex in a non-generic way. The reconstruction of the two-loop contributions as described above renders these approximations unnecessary. 

\begin{figure}
\begin{center} 
\ifbuidln\includegraphics[width=.6\textwidth]{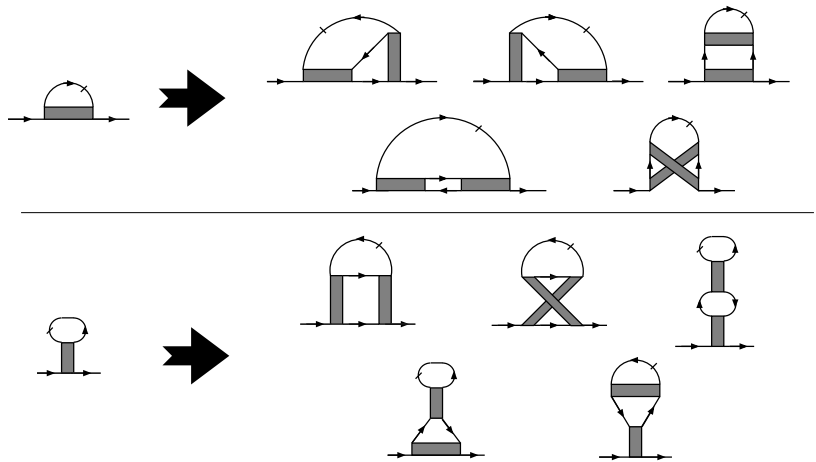}\fi
\end{center} 
\caption{Second order diagrams for the self energy. The diagrams correspond either to
tadpole$\times$one-loop contributions or real two-loop terms. We only keep the
logarithmically divergent two-loop diagrams. One internal line is at scale $\Lambda$,
indicated by the bar in Fig. \ref{twoloopdias}, and one of the other two lines
corresponds to a single-scale propagator at scale $\Lambda'$ that is being integrated
over. The  third line is an interacting Green function with support on modes above
$\Lambda'$.}
\label{twoloopdias}
\end{figure}

There are two distinct one-loop diagrams for the flow of the self energy 
(the Hartree and the Fock diagram) 
and 5 one-loop diagrams (one particle-particle and four particle-hole diagrams) 
for the coupling function (see Figures 4 and 5 in Ref.\ \onlinecite{sh}). 
Therefore we get ten diagrams from inserting the flow of the coupling function 
into the two one-loop diagrams for $\Sigma_\Lambda$, shown in Fig. \ref{twoloopdias}. 
Four of these diagrams are of one-loop$\times$tadpole-type, hence frequency-independent
in our approximation and thus do not contribute to $Z_\Lambda$.
Thus we are left with six two-loop diagrams for 
$(d/d\Lambda) \, \partial_\omega \Sigma_\Lambda (0, \vec{k})$.

In condensed notation the RG equations for the selfenergy are
\begin{equation}
\frac{\partial \Sigma_\Lambda}{\partial\Lambda} 
= 
\gamma^{(4)}_\Lambda * S_\Lambda
= 
\left(
\int d\Lambda' \;
V_{\Lambda'} *
S_{\Lambda'}* G_{\Lambda'}* 
V_{\Lambda'} \right) * S_\Lambda 
\end{equation}
Here the star stands for the momentum integration and Matsubara frequency 
summation according to the standard diagrammatic rules 
(for details, see Refs.\ \onlinecite{hsfr,sh}).
Similarly, the equation for the coupling function reads
\begin{equation}
\frac{\partial V_\Lambda }{\partial\Lambda} 
= 
V_{\Lambda}  * 
S_{\Lambda} * 
G_{\Lambda} *\, 
V_{\Lambda} .
\end{equation}    
In the following we consider the renormalized vertex and coupling function
\begin{eqnarray} 
\renG^{(4)}_{\Lambda,s_1s_2s_3s_4} (\vec{k}_1,\vec{k}_2,\vec{k}_3,\vec{k}_4) 
&=& 
\left[ 
Z_\Lambda (\vec{k}_{1}) Z_\Lambda (\vec{k}_{2}) Z_\Lambda (\vec{k}_{3}) Z_\Lambda
(\vec{k}_{4})
\right]^{1/2} 
\gamma^{(4)}_{\Lambda,s_1s_2s_3s_4} (\vec{k}_1,\vec{k}_2,\vec{k}_3,\vec{k}_4) 
\label{gammaren}
\\ 
\renV_\Lambda (\vec{k}_1,\vec{k}_2,\vec{k}_3) 
&=& 
\left[ 
Z_\Lambda (\vec{k}_{1}) Z_\Lambda (\vec{k}_{2}) Z_\Lambda (\vec{k}_{3}) Z_\Lambda
(\vec{k}_{4})
\right]^{1/2} \, 
V_\Lambda(\vec{k}_1,\vec{k}_2,\vec{k}_3) 
\label{renV}
\end{eqnarray}
They describe the interaction between the quasiparticles 
with renormalized amplitude $Z_\Lambda$ at the corresponding energy scale. 
If the limit $\Lambda \to 0$ can be taken and $Z_\Lambda$ remains nonzero, 
the quasiparticle scattering amplitudes of the emergent Landau Fermi liquid 
can be read off eq.\ (\ref{renV}). 

The flow equations can now be expressed in terms of $\renG$ and $Z_\Lambda$ 
and the $C_\Lambda$ defined in (\ref{Zpropagators}) as
\begin{equation}\label{Zflo}
\frac{\dot Z_\Lambda}{Z_\Lambda} (\vec{k}_F) = 
- i \partial_\omega (\renG * \dot C_\Lambda) (0,\vec{k}_F)
\end{equation} 
(here the dot denotes $\Lambda \partial/\partial \Lambda$) and
\begin{equation}\label{renGflo}
\dot\renG (k_1,k_2,k_3,k_4) = 
(\renG * C_\Lambda * \dot C_\Lambda * \renG)  (k_1,k_2,k_3,k_4)
+ 
\sum_{i=1}^4 \frac{\dot Z_\Lambda (\vec{k_i})}{ 2 Z_\Lambda (\vec{k_i})}
\;
\renG  (k_1,k_2,k_3,k_4)
\end{equation} 
In an expansion in terms of the bare coupling function, 
the last term on the right hand side of (\ref{renGflo})
is of third order. In the Luttinger model, this term cancels exactly 
against the third order contribution to the coupling function flow\cite{BGPS,busche}.
Because of the cancellation between the particle--hole and particle--particle terms
in the first term, the coupling in the Luttinger model does not flow in this
approximation,
as it must be. In higher dimensions we also have to drop this term 
to remain consistent to second order. Keeping this term would require also taking 
into account the higher orders in the flow of the coupling function,
that is, truncating only at $\gamma^{(8)}$. 

Therefore, in the following we drop the second term in (\ref{renGflo}). 
As just discussed, this is consistent both in 1D and in higher dimension
and thus allows us to compare the flow in the Luttinger model
with that in weakly and fully two--dimensional situations.
An important feature of the equations is that the $Z$ factors 
do not appear in the equation for the renormalized four point vertex any more, 
so that the flow of the renormalized four point function is the same 
as that of the (unrenormalized) four point function when the $Z$ factor is dropped
altogether. Therefore a small $Z$ factor can only prevent a flow to strong coupling 
through (\ref{gammaren}). 
One conclusion of this work will be that - in contrast with the 1D case - 
in the 2D system the suppression of the quasiparticle weight is weak 
as long as the coupling function remains in the perturbative range. 
The full two-loop flow of the coupling function is left for future work,
but the value of $Z$ obtained from our calculation will already give us 
a first estimate of the size of these two--loop terms.

As discussed above, $\renG$ is now replaced by the integrated equation 
(\ref{renGflo}), with $\renV$ replacing $\renG$ on the right hand side,
in (\ref{Zflo}). We can now solve for $Z$ in the one--dimensional case.
If $\renV$ is independent of scale and has initial value $U$, 
eq.\ (\ref{Zflo}) reads (in the local approximation described above)
\begin{equation}
 \frac{\dot Z}{Z} =  U^2 \; \alpha 
\end{equation}
with $\alpha$ the value of the two--loop 
contribution and $U$ the coupling constant. This integrates to the power law
\begin{equation}
Z = \left(\frac{\Lambda}{\Lambda_0}\right)^{\alpha U^2}
\end{equation}
It follows that the two--point function has an anomalous decay exponent given by
$\alpha U^2$. 
This argument also shows that it is not necessary to include a rescaling of the momenta
in the RG transformation to obtain anomalous dimensions, contrary to some
claims\cite{kopietz}.  

In higher dimensions and for curved Fermi surfaces, the overlapping loop 
effect\cite{FST} implies that the two--loop integral is small, so that
(\ref{Zflo}) reads 
\begin{equation}\label{Zdge2}
\frac{\dot Z}{Z} = \tilde\alpha U^2 \; 
\cases{  \Lambda \log \Lambda & $d=2$ \cr \Lambda & $d \ge 3$,}
\end{equation}
the solution of which remains close to one for all $\Lambda \ge 0$ if $U$ is small. 
However, in that case, the coupling function is not constant because of the 
Cooper instability, so that eventually $Z$ would also differ significantly 
from $1$. On the other hand, to get a bound we only need to replace $U^2$ 
by the maximal value of the $\Lambda$--dependent coupling function in (\ref{Zdge2}). 
This shows that $Z$ does not differ much from $1$ before the couplings get of
order $ 1 /\sqrt{\Lambda}$, which is outside of the range where the one--loop coupling
flow 
is justified\cite{sh}. So $Z$ plays no important role in the flow. 
Moreover, when restricting to 
temperatures above the critical temperature for superconductivity,
the couplings remain small if they started out small, 
and the $Z$ factor remains close to $1$: 
the system is a weakly coupled Fermi liquid
\cite{salmcrg,DR}.

When the Fermi surface contains van Hove points or flat parts,
the $Z$ factor can, however, get strongly suppressed. We investigate this in Sec.\ V.

\section{The Hubbard chain}
In this section we apply our formalism to the one-dimensional Hubbard chain. 
We have already seen that when the couplings do not flow, the $Z$ factor 
obeys a power law, so that our approximation does describe the 
Luttinger exponents. Thus the numerical calculations in this section
 serve mainly as a check for the numerical implementation 
but they also gives some insight on the magnitude of band structure effects. 
The tight binding dispersion along the chain is given by
\begin{equation} \label{xhop}
\epsilon_{\vec{k}} = -2t \, \cos k_x - \mu \, ,  
\end{equation}
where the chemical potential $\mu$ determines the particle number. 
In the usual $g$-ology terminology\cite{bourbonnais} 
one considers four coupling constants $g_1, \dots, g_4$ 
with two incoming and one outgoing quasiparticle on the Fermi surface. 
For the Hubbard interaction $U \sum_{i} n_{i,\uparrow} n_{i,\downarrow}$ 
at the initial stage of the RG procedure they all have the same value $g_i =U$. 
First let us consider the case away from half filling. 
In this case the one-loop flow of the interactions does not lead to strong coupling. 
The coupling $g_3$ involves scattering into states away from the Fermi surface 
and does not influence the low energy physics. 
$g_4$ does not couple to any logarithmically diverging channel 
and does not flow in the usual RG schemes. 
$g_1$ is irrelevant and is driven to zero towards lower scales. 
$g_2$ is somewhat reduced as the difference $2g_2 -g_1$ is a constant in the one-loop
flow. 
The simplest model completely neglects the flow of $g_2$ and sets $g_1=0$. 
As discussed in the preceding section, and in accordance with 
Refs. \onlinecite{metzner1D} and \onlinecite{bourbonnais}, 
in the local approximation (described in the previous section) to the RG equations the quasiparticle weight 
is suppressed like
\begin{equation} \label{g2z}
Z_\Lambda 
\sim \left( \frac{\Lambda}{\Lambda_0} \right)^{g_2^2/(4 \pi^2 v_F^2)} 
\, .  
\end{equation}
If we allow the coupling constants $g_1$ and $g_2$ to flow, 
this gets modified\cite{bourbonnais} to 
\begin{equation} \label{g1g2z}
Z_\Lambda \sim 
\left( \frac{\Lambda}{\Lambda_0} \right)^{(2g_2-g_1)^2/(16 \pi^2 v_F^2)} 
\exp\left[ 
- \frac{3}{16} \int_{\Lambda_0}^\Lambda d\Lambda' \frac{g_{1,\Lambda'}^2}{\Lambda'} 
\right]  
\, .  
\end{equation}
\begin{figure}
\begin{center} 
\ifbuidln\includegraphics[width=.7\textwidth]{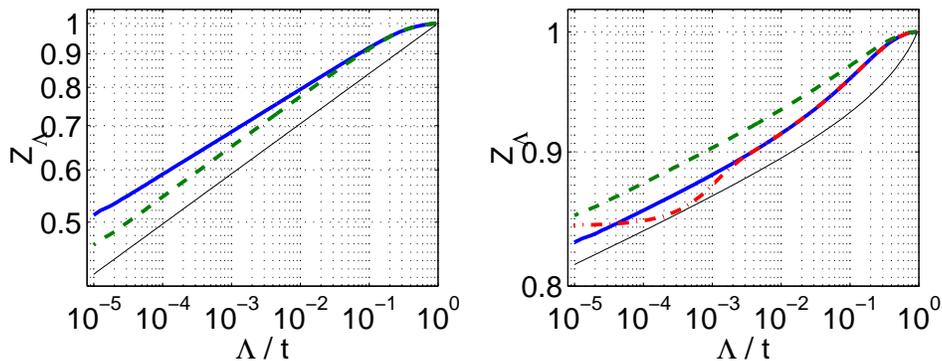}\fi
\end{center} 
\caption{Flow of the quasiparticle weight in the Hubbard chain 
when the flow of the interactions is suppressed (left plot) or allowed (right plot). 
The thin solid line shows the analytical expectations described in the text. 
The thick  line show the numerical results at zero temperature 
in the $\Lambda$-local approximation to the RG equations, 
while the thick solid lines are without the local approximation. 
The thick dashed-dotted line is for $T=10^{-3}t$. 
All data are for $\mu = -t$ and $U=3t$.}
\label{lumo}
\end{figure} 
In Fig. \ref{lumo} we compare these expressions with our numerical RG scheme. 
If we choose the same local approximation the agreement is excellent 
down to scales $10^{-5} t$ (we start the flow at $\Lambda_0 =t$). 
We also observe some deviations in the initial flow 
due to the non-linearity of the dispersion relation. 
Below $\Lambda \approx 10^{-2}t$ these deviations are negligible 
and could be absorbed in the powerlaws (\ref{g2z}) and (\ref{g1g2z}) 
by simple prefactors.
The RG equations that are nonlocal in the scale yield lead to quantitative corrections.
If we neglect the flow of the coupling function (left plot in Fig. \ref{lumo}), 
the non-local corrections slightly weaken the Luttinger exponent. 
In the case where the coupling function is allowed to flow (right plot), 
the non-local corrections suppress the $Z$ factors further 
because the couplings $g_1$ and $g_2$ are decreasing functions of the RG scale. 
Nonzero temperature $T$ causes the expected saturation of the flow of $Z$ 
at scales $\Lambda < T$.
Note that in our case, where $U=3t$, corresponding to $g_2^2/(4 \pi^2 v_F^2) \approx
0.076$, 
and the initial scale $\Lambda_0=t$, the suppression of $Z$ is relatively weak.

\section{Effects of interchain hopping}
Next let us consider  the case of nonzero hopping $t_\perp>0$ between Hubbard chains.
This leads to the dispersion relation
\begin{equation} \epsilon_{\vec{k}} = -2t \, \cos k_x - 2t_\perp \cos k_y - \mu \, .
\label{xyhop} \end{equation} 
The curvature of the Fermi surface (see Fig. \ref{tp2plot}) induced by $t_\perp$
prevents the cancellation of the one-loop flow of the interaction and cuts off the
divergence of the self energy derivatives. 
Phase space considerations indicate
that above this cutoff scale $\Lambda_{\perp}$ set by the curvature
the system will resemble a one-dimensional system, while below $\Lambda_{\perp}$ it
will behave like a two-dimensional Fermi liquid and maybe undergo a flow to strong
coupling.  Analytic expressions for the quasiparticle weight in coupled chains and the
crossover from 1D to 2D behavior can be found in the review by
Bourbonnais\cite{bourbonnais}. They are based on a random phase approximation for the
interchain motion and state that 
\begin{equation}   \Lambda_{\perp} \sim t_\perp \,  \left(
\frac{t_\perp}{\mathrm{bandwidth}} \right)^{\frac{\theta}{1-\theta}} \label{tperpren}
\, , \end{equation} 
where the second factor describes the effective decrease of the interchain hopping by
the renormalization of the quasiparticles on the chains. $\theta$ is the exponent that
describes the decay of $Z_\Lambda$ on a single chain. In our case $\theta \ll 1$. Thus
we expect $\Lambda_\perp \approx t_\perp$.

\begin{figure}
\begin{center} 
\ifbuidln\includegraphics[width=.45\textwidth]{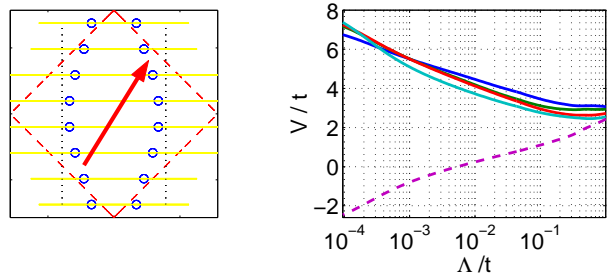}\fi
\end{center} 
\caption{Left: Fermi surface for interchain hopping $0.3t$ and the 16 FS points used in
the $N$-patch calculation. Right: Flow of the most repulsive (solid line) and most
attractive (dashed line) coupling constants. The repulsive scattering processes are
largest between the almost nested flat sides of the Fermi surface (see arrow in left
plot). The temperature is $T=10^{-4}t$, initial $U=3t$. The minimal quasiparticle weight at $\Lambda =
T$ is 0.89. }
\label{g20tp03}
\end{figure}  
In absence of interchain hopping and Umklapp scattering, the one-loop flow of the
coupling constants does not lead to strong coupling because the Cooper diagrams cancel
the $2k_F$-particle-hole diagrams that would otherwise lead to a nesting instability.
This delicate balance gets destroyed by nonzero $t_\perp$ as particle-particle and
particle-hole diagrams depend on the transverse wavevector in a different way. Partial
nesting between almost parallel parts of the Fermi surface generates a growth of some
coupling constants (see Fig. \ref{g20tp03}). Interactions between different layers of
coupled chains can in principle induce spin density wave ordering with the
corresponding wavevectors at finite temperature. With increasing $t_\perp \approx 0.5t$
and larger the Fermi surface becomes more and more rounded and sufficiently away from
half filling the nesting features get washed out.

The specific question we are interested in is whether the flow of the quasiparticle
weight becomes significant in the scale range where these nesting features develop.
This is important as many studies of potential instabilities in quasi-1D materials
neglect possible selfenergy effects. Here we only consider the change of the
quasiparticle weight, but a thorough analysis would also include the renormalization of
the band dispersion and lifetime effects.

In Fig. \ref{tp2plot} we compare the flow of the quasiparticle weight for three
different values of the interchain hopping $t_\perp$. The anisotropy of the $Z$-factor
along the Fermi surface is very weak and can be neglected for these values of $t'$.
Even for case of $t_\perp =0$ the $Z$-factor is suppressed only weakly down to scales
of $10^{-4}t$ where the flow is cut off by the nonzero temperature\footnote{We refrain
from going to lower scales as the numerics for many patches becomes very slow below
these scales.}. 
For $t_\perp=0.002t$ and $t_\perp=0.005t$ the suppression of $Z$ is even weaker and
gets cut off at scales $\sim t_\perp$. Therefore, at least for the parameters
considered here, the flow of the quasiparticle weight does not qualitatively alter the
conclusions drawn from the one-loop flow of the coupling function. We expect that in
most cases where the coupling function grows large at low scales the flow of $Z$ can be
neglected.  In order to destroy the quasiparticle effectively one has to go to
extremely low energy scales and temperatures. On the other hand, the suppression of $Z$
towards lower energies may still be visible in tunneling or optical measurements. 

\begin{figure}
\begin{center} 
\ifbuidln\includegraphics[width=.4\textwidth]{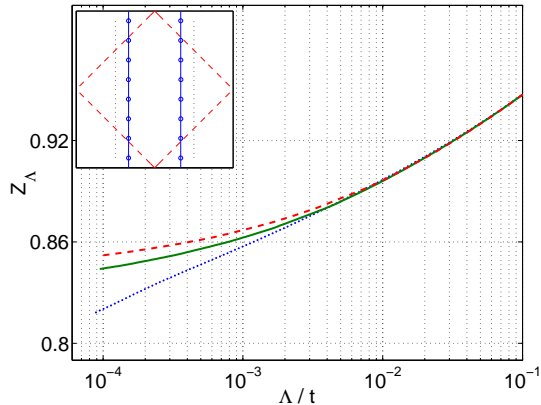}\fi
\end{center} 
\caption{Flow of the quasiparticle weight in the Hubbard chain coupled by a interchain
hoppings $t_\perp = 0$ (dotted line), $0.002t$ (solid lines) and $0.004t$ (dashed
line). The anisotropy of the quasiparticle weight over the Fermi surface is below the
numerical precision. The temperature is $T=10^{-4}t$, initial $U=3t$.}
\label{tp2plot}
\end{figure} 

\section{The quasiparticle weight in two dimensions}
Next we study the 2D Hubbard model close to half filling. 
The one-loop flow of the coupling function has been analyzed extensively 
for zero\cite{zanchi,binz} and nonzero values\cite{halboth,hsfr} 
of the next-nearest neighbor hopping parameter $t'$. 
As discussed earlier, in our approximation 
the flow of the renormalized coupling function is unmodified 
compared to the earlier studies neglecting the flow of the quasiparticle weight. 
Here we focus on the case $t'=0$ and on the flow of the quasiparticle weight. 
We want to analyze a) how strong the suppression of the quasiparticle weight is 
and whether these two-loop effects get comparable to the one-loop terms 
that drive the coupling function; 
b) whether the $Z$-factor develops a distinctive anisotropy around the Fermi surface, 
e.g. at higher temperature, when the flow of the coupling function remains finite.

\begin{figure}
\begin{center} 
\ifbuidln\includegraphics[width=.7\textwidth]{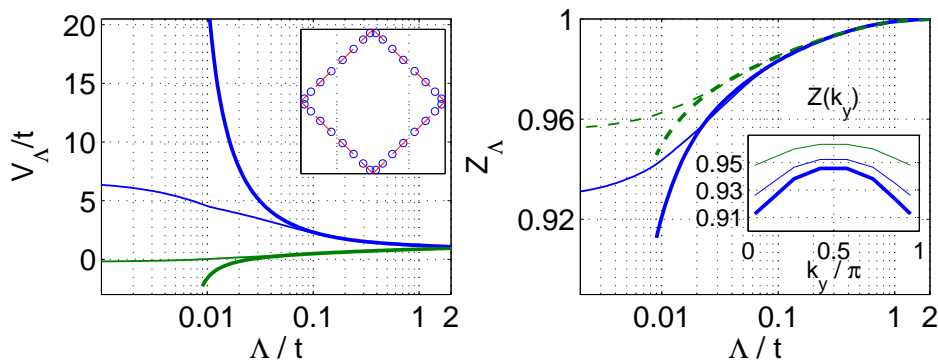}\fi
\end{center} 
\caption{Left: Flow of maximal and minimal coupling constants for the almost
half-filled $t'=0$ Fermi surface (see inset) with chemical potential $\mu=-0.01t$, initial $U=t$ and $N=32$ points, at
$T=0.001t$ (thick lines) and $T=0.03t$ (thin lines). 
Right: Flow of the quasiparticle weight close to the
saddle point (solid line) and in the BZ diagonal (dashed line)  at
$T=0.001t$ (thick lines) and $T=0.05t$ (thin lines). The inset shows the
small variations of $Z$ around the FS for $T=0.001t$ at the scale 
where the largest coupling constant is $10t$ (upper line), $20t$ and  
$30t$ (bottom line), i.e. already far out of the perturbative range. } 
\label{g32mu0}
\end{figure}

\begin{figure}
\begin{center} 
\ifbuidln\includegraphics[width=.7\textwidth]{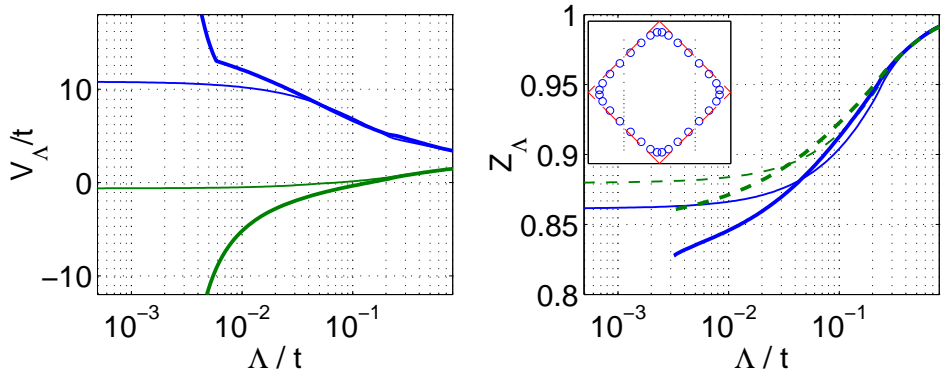}\fi
\end{center} 
\caption{Left: Flow of maximal and minimal coupling constants for the $t'=0$ Fermi
surface with electron density $0.87$ per site, initial $U=2.25t$ and $N=32$ points at
$T=0.001t$ (thick lines) and $T=0.2t$ (thin lines). Right: Flow of the quasiparticle
weight close to the saddle points (solid line) and in the BZ diagonal (dashed line).
Again thin lines are for $T=0.2t$ and thick lines for $T=0.001t$. The inset shows the
32 point on the Fermi surface used in the $N$-patch calculation.} 
\label{w32mu23}
\end{figure}

First let us describe the $N$-patch results shown in Fig. \ref{g32mu0} for half filling 
and in Fig. \ref{w32mu23} for electron density $\langle n \rangle = 0.9$ per site. 
We observe that the quasiparticle weight gets increasingly suppressed toward low scales
where the coupling functions get large. The suppression is strongest close to the van
Hove points. 
Thus the flow is consistent with a vanishing quasiparticle weight at low energies 
at least in the saddle point regions.
Nevertheless we note that even if we follow the growth of the renormalized coupling
function 
far out of the perturbative range, the suppression of $Z$ remains rather weak. 
This shows that the two-loop effects diminishing the quasiparticle weight are small 
compared to the one-loop contributions driving the coupling function 
and that the one-loop approximation\cite{zanchi,halboth,hsfr} for the latter 
may be justified in this scale range.    

A similar picture is found away from half filling. 
In Fig. \ref{w32mu23} we show data for $\langle n \rangle = 0.87$ and initial
$U=2.25t$. 
At low temperatures the flow goes towards a $d$-wave superconducting instability. 
The quasiparticle weight gets somewhat suppressed towards the instability, 
but the renormalized coupling constants flow off to strong coupling. 
Although the flow of the $Z$-factors is driven by the interactions, 
it still continues an approximately logarithmic decrease 
while the interactions seem to diverge with some power-law. 
This indicates that at these scales the two-loop contributions are much weaker 
than the one-loop terms that drive the flow to strong coupling. 
At higher temperatures, $T\ge 0.2t$ the coupling function does not diverge, 
and we can follow the flow to lowest scales. 
The flow of the quasiparticle weight gets cut off at scales $\Lambda \approx T$ 
and remains close to unity over the full Fermi surface.
Like in the quasi-1D cases studied in the last section 
the suppression of $Z$ is only mildly anisotropic.

Next we interpret these findings and put them into context with other works in the literature.
The flow of the quasiparticle weight close to the saddle points was studied 
by Dzyaloshinski\cite{dzyaloshinski} for $t'\not= 0$ in a system restricted to 
the saddle point regions, and more recently by Zanchi\cite{zanchiZ} 
for the half-filled case using a $N$-patch formalism related to ours. 
Dzyaloshinski\cite{dzyaloshinski} neglected the logarithmically divergent particle-hole
diagrams 
with respect to the logarithm-squared particle-particle diagrams. 
Consequently in the repulsive Hubbard model his flow does not lead to strong coupling. 
Nonetheless the quasiparticle weight approaches zero at low scales, 
similar to the Luttinger liquid case. However, at least for our choice of parameters, 
the particle-hole diagrams always drive the flow to strong coupling, 
even if the nesting is weaker. 
Zanchi\cite{zanchiZ} argued that for the half-filled case and $t'=0$ 
the quasiparticle weight approaches zero at the saddle points at scales 
above the divergence scales of any susceptibilities, indicative of an anomalous normal state without long range order.
The suppression of the quasiparticle weight is in qualitative agreement with our
results. However we note that in the perturbative range, where the approach is valid, 
the suppression of $Z$ is only weak. Therefore in our opinion no strong conclusions 
regarding the order of closely competing poles in the flow of $Z$ and various susceptibilities can be drawn. 
In view of the weakness of the two-loop effects it appears to be more likely that for the half-filled perfectly nested case the barely attenuated growth of the coupling function leads to 
spin-density wave (SDW) order in the ground state where quasiparticles exist 
only above the SDW energy gap or at temperatures above the instability. 
This is corroborated by the fact that at temperatures above the runaway flow of the interactions (see thin lines in Fig. \ref{g32mu0}) the flow of the quasiparticle weight saturates at low scales and remains nonzero. Thus, although we cannot rule out the possibility, we do not find any evidence for a non-Fermi liquid or Luttinger-liquid like state at temperatures above the flow to strong coupling.  

A similar expectation regarding the one-particle spectrum 
holds for densities less than half filling: 
the growth of the renormalized coupling function out of the perturbative range 
indicates the opening of a gap at least on parts of the Fermi surface. 
In the case $\langle n \rangle = 0.9$ per site we expect 
superconducting long range order of $d_{x^2-y^2}$-symmetry. In our approximation the flow of the interactions is unchanged by the flow of the quasiparticle weight and develops a strongly dominant diverging pair scattering in the $d_{x^2-y^2}$-channel as discussed by several authors\cite{zanchi,halboth,hsfr,tsai,binz}. 
The weak suppression of the quasiparticle weight down to scales where the pair scattering is already much larger than the bandwidth shows the irrelevance of two-loop effects. 
Thus our results affirm the evidence for $d$-wave superconductivity in the Hubbard model at weak to moderate $U$.
 
Note however that there are examples of one-dimensional systems 
where a one-loop flow to strong coupling does not lead to 
a (quasi-)long-range ordered ground state. In the half filled two-leg Hubbard ladder \cite{lin} 
the single particle spectrum is fully gapped, yet all spin- and charge correlation functions remain exponentially decaying. 
We pointed out\cite{hsfr} that in the 2D Hubbard model for nonzero $t'$ 
the flow to strong coupling strongly resembles the one in the two-leg ladder 
and that this may indicate a novel strong coupling state 
for a certain parameter range of the 2D Hubbard model different from the $t'=0$ case studied above.    

\section{Conclusions and outlook}
The $N$-patch renormalization group scheme enables us to trace the change in behavior
of weakly correlated electron systems as a function of dimension, by interpolating the
ratio of the hopping parameters in the two directions from zero to one. 
It reproduces the Luttinger liquid behavior in the one-dimensional models 
away from commensurate band filling on one side 
and the Fermi liquid state with its various infrared instabilities on the other side. 
The present RG calculation still involves a cascade of approximations 
but our results match well with established knowledge acquired by other methods. 
In one spatial dimension the quantitative precision of the RG scheme 
can be checked directly by exact numerical techniques: 
in the study of boundaries of Luttinger liquids 
the comparison with  density matrix renormalization group calculations 
turned out to rather promising\cite{meden}. 
Biermann et al.\cite{biermann} used extended dynamical mean-field theory (DMFT) 
to calculate the quasiparticle weight in coupled chains. 
Their findings for nonzero interchain hopping are qualitatively similar to the RG
results, 
e.g. the suppression and also the anisotropy  of the $Z$-factors 
are relatively weak along the Fermi surface 
(note however that the RG calculation for $Z$ applies only at weak couplings, 
and higher order corrections might further decrease the $Z$-factors).
We believe that the true strength of the $N$-patch RG lies in its flexibility and
transparency. 
Due to these properties the method may open the way to an unified qualitative picture 
of low dimensional weakly coupled electron systems. 

A direct conclusion of this work is the partial justification 
of the neglect of the wave function renormalization 
in several previous renormalization group studies of the 2D Hubbard
model\cite{zanchi,halboth,hsfr,tsai}. 
At least for the parameters used in these studies 
and over the scale range where the perturbative scheme is expected to be valid, 
the quasiparticle weight remains nonzero 
and its flow does not affect the one-loop flow to strong coupling 
in the second order approximation. 
Similar statements can be made about the one-loop Fermi surface shift\cite{hsfr} 
and the quasiparticle scattering rate\cite{honedhd}. 
The physical picture that is suggested by these findings for the Hubbard model 
close to half filling is that the most plausible interpretation 
of the runaway flow of the coupling function is the formation of an energy gap 
at least on parts of the Fermi surface and not a gapless state resembling a
Luttinger liquid. Due to the limited validity range of the weak coupling approach 
our results can not support a conjectured pseudogap\cite{zanchiZ} at temperatures 
above the antiferromagnetic instability in the half-filled Hubbard model. 
On the other hand our calculations show that the antiferromagnetic instability and also the $d$-wave superconducting instability found in earlier studies \cite{zanchi,halboth,hsfr,tsai} persist if the flow of the quasiparticle weight is included. In particular the possibility of $d$-wave superconductivity in the Hubbard model is still debated\cite{laughlin,maier,sorella}, however mostly at larger $U$. Our analysis underpins the positive answer for the  weak coupling sector of the model.

An analysis of the renormalization of the dispersion relation 
and of the higher--order flow of the coupling function 
will provide further checks for the existing results.  
Regarding  further applications of the method 
we mention the wide class of quasi-1D conductors like Bechgaard or Fabre
salts\cite{bourbonnais2}. 
These systems show a variety of magnetically ordered or charge-ordered states 
and many undergo transitions into supposedly unconventional superconducting states. 
The hopping parameters of these materials are known to a good extent\cite{jerome} and a weak coupling description is expected to make sense. 
Thus our method should be well applicable and useful 
to understand the main features of the phase diagrams of these materials 
and the interplay between one- and higher-dimensional physics. 
Some steps in this direction have been undertaken by Duprat and
Bourbonnais\cite{duprat}.  

We have used an approximation scheme where all relevant contributions 
up to second order in the interactions are taken into account. 
This correctly reproduces the Luttinger liquid case, 
where the flow of the renormalized interaction vanishes also in second
order\cite{BGPS,busche}. 
In those cases where the second order flow leads to strong coupling, 
higher order contributions will become large at some scale.
The physical content of the strong coupling fixed points 
in two-loop $g$-ology studies\cite{solyom,bourbonnais} is, however, rather
questionable, 
as the fixed point coupling constants are typically larger than the bandwidth 
and thus outside the perturbative range. 
Moreover the information obtained by careful one-loop studies of 1D systems 
agrees well with the results of numerical techniques or bosonization. 
Of course, our RG method can in principle be used later on to investigate 
the higher order flow of the interactions as well. 
For the time being, note that our present scheme at least provides estimates 
when higher order effects such as the flow of the quasiparticle weight become
important. 

In addition we note that the $N$-patch approach in its present form 
can be used to extract a precise Landau-Fermi liquid description 
of two-dimensional systems like the Hubbard model at temperatures 
above the flow to strong coupling, extending existing studies\cite{fuseya,frigeri} 
that do not consider the renormalization of the quasiparticle weight. 
In view of the somewhat harmless flow of the $Z$-factors 
in the two-dimensional models discussed above 
we do, however, not expect drastic changes of the results. 

\section*{Acknowledgements} 
We thank Patrick Lee, Walter Metzner, Maurice Rice and Ashvin Vishwanath for valuable
discussions. This research was supported in part by the National Science Foundation under Grant No. PHY99-07949. C.H. acknowledges financial support by the DFG (Deutsche Forschungsgemeinschaft).

\section*{Appendix A: Sharp cutoff limit}
We consider a general class of cutoff functions given as
\begin{equation}
\chi_\Lambda (\vec{p}) = K(\epsilon_{\vec{p}} - \Lambda) + K(-\epsilon_{\vec{p}}
-\Lambda)
\end{equation}
where $K$ is a smooth function that increases from $0$ to $1$ in a small neighborhood 
of size $\eta > 0$ of zero. 
$K$ is constructed as follows. Let $h$ be a smooth function of $x$ that vanishes for 
$|x| > 1$ with the properties $h(x) \ge 0$ for all $x$ and $\int_{-\infty}^\infty h(x)
dx = 1$.
Define $H(x) = \int_{-\infty}^x h(u) du$ and let $K(x) = H(x/\eta)$. 
As $\eta \to 0$, $K$ approaches the Heaviside function: $\lim_{\eta\to 0} K(x) = \Theta
(x)$,
and thus becomes a "sharp" cutoff function. 
We note that we have to take the limit $\eta \to 0$ or else use this as a cutoff
function
only for $\Lambda > \eta$ because for $\Lambda < \eta$, $K$ no longer vanishes near
zero.
The following argument for taking the limit $\eta \to 0$
applies with trivial modifications also to the often taken choice 
of a $\chi_\Lambda (E) $ that depends only on the ratio $E/\Lambda$. 

In the following we show that in the sharp cutoff limit, the propagators 
appearing in the RG flow equation are given by (\ref{Galapeno}) 
and (\ref{Salapeno}). Moreover, since the single scale propagator contains
$\delta (\pm \epsilon_{\vec{p}} - \Lambda)$ and the full propagator 
contains $\theta (\pm \epsilon_{\vec{p}} - \Lambda )$,
formal expressions like $\delta(x) \theta(x)$ appear
and we determine what they really mean in the limit. 
As it turns out, the correct rule is that one has to replace the 
cutoff function by a variable and integrate it from $0$ to $1$. 
We discuss the integral
\begin{equation}
I_\pm (q) = \int dp\; 
G_\Lambda (q \pm p) S_\Lambda (p)\; \Phi_\Lambda (p,q)
\end{equation}
which is prototypical for the integrals appearing in the flow equation. 
The other cases appearing are that where $G_\Lambda$ is replaced by $1$
(in the equation for the selfenergy) and when $G_\Lambda$ is replaced by a 
product of $G_\Lambda$'s (such terms appear in the flow equations for the 
higher $m$--point functions. The function $\Phi_\Lambda$ is a product of 
vertex functions. 

Let us first consider the case without van Hove singularity. 
In this case, we can introduce a radial coordinate $E= \epsilon_{\vec{p}}$ and
an angular coordinate $\theta$, so that $\vec{p} = \vec{p}(E,\theta)$ is now 
a function of $E$ and $\theta$. We also write $p(E,\theta)$ for $(p_0,\vec{p}
(E,\theta))$.
We have
$I_\pm (q) = T \sum_{p_0} \int d\theta\; (X_+ + X_-)$ with 
\begin{equation}
X_+= \int_0^\infty dE \; J(E,\theta)\; S_\Lambda (p) G_\Lambda (q \pm p) 
\;
\Phi (p,q)
\end{equation}
and $X_-$ the integral over negative $E$.  Because we restricted to positive $E$,
$\chi_\Lambda (\vec{p})$ simply becomes $K(E - \Lambda) $. Inserting the definition 
of $K$ in terms of $H$ and changing variable from $E$ to $u=(E-\Lambda)/\eta$, 
we get
\begin{equation}
X_+= \int_{-\Lambda/\eta}^\infty du \; 
J(\Lambda+u\eta,\theta)\;
\frac{-H'(u)
(ip_0-\Lambda-u\eta)}{(ip_0-\Lambda-u\eta-H(u)\Sigma_\Lambda(p(\Lambda+u\eta,\theta))^2}
\;
G_\Lambda(q \pm p )
\;
\Phi_\Lambda (p,q)
\end{equation}
Here $J$ is the Jacobian of the change of variables; at fixed $E$,
its integral over $\theta$ gives the density of states at $E$,
and we inserted a $\theta$ function to restrict the integration to
$u \ge - \Lambda/\eta$. Because we are at a positive temperature,
the integrand is bounded by an integrable function. Moreover its 
limit as $\eta \to 0$ exists pointwise. Thus, by dominated convergence,
the limit $\eta \to 0$ exists and can be taken inside the integral, so that  
\begin{equation}
X_+= \int_{-\infty}^\infty du \;
J(\Lambda,\theta)\;
\frac{-H'(u) (ip_0-\Lambda)}{(ip_0-\Lambda-H(u)\Sigma_\Lambda(p(\Lambda,\theta)))^2}
\;
G_\Lambda(q \pm p )
\;
\Phi_\Lambda (p,q)
\end{equation}
Note that the limit of $G_\Lambda$ is not a continuous function,
but this does not matter for the exchange of limits. 
The function $\Phi_\Lambda$ also has a limit as $\eta  \to 0$. 
This follows by the argument we just outlined and by iteratively solving the 
differential equation. 
Let $q \ne 0$. We can change 
variables to $H$ and are left with
\begin{equation}
X_+=
J(\Lambda,\theta)
\;
G_\Lambda(q \pm p(\Lambda,\theta) )
\;
\Phi_\Lambda (p,q)
\;
\int_{0}^1 dH \;
\frac{-(ip_0-\Lambda)}{(ip_0-\Lambda-H\Sigma_\Lambda(p(\Lambda,\theta)))^2}
\end{equation}
which corresponds to the statement made above that the cutoff function has to be
replaced by a variable $H$ that gets integrated from $0$ to $1$. 
The integral gives $(ip_0-\Lambda-\Sigma_\Lambda(\Lambda,\theta))^{-1}$
so the final integration over $\theta$ and $p_0$ corresponds to a $p$--integral 
with a propagator $S_\Lambda^{(+)} (p) =
\delta (\epsilon_{\vec{p}} - \Lambda)\;(ip_0-\epsilon_{\vec{p}}-\Sigma_\Lambda(p))^{-1}
$.
The contribution from $X_-$ is, of course, similar, with $\delta(\varepsilon_{\vec{p}}
+ \Lambda)$
replacing $\delta(\varepsilon_{\vec{p}} - \Lambda)$.

For the special case $q=0$ the energies and cutoff functions in $G_\Lambda$ 
are identical to those in $S_\Lambda$, so that 
\begin{equation}
X_+=
J(\Lambda,\theta)
\;
\Phi (p,q)
\;
\int_{0}^1 dH \;
\frac{-(ip_0-\Lambda)\; H}{(ip_0-\Lambda-H\Sigma_\Lambda(p_0,p(\Lambda,\theta)))^3}
\end{equation}
The integral over $H$ now gives $- {1\over 2} (ip_0-\Lambda - \Sigma_\Lambda)^{-2}$.
Note the extra factor $1/2$. 
Let us compare this with the result of taking $q \to 0$ in the above results.
If we integrate over $\theta$, set $q_0=0$ and take the limit $q \to 0$,
we also get only an integral over "half of" the Fermi surface:
the intersection of the level set 
$L_\Lambda = \{\vec{p}:\epsilon_{\vec{p}}=\Lambda\}$ 
(along which we integrate)
with the support of $\theta (\epsilon_{\vec{p+q}} - \Lambda)$ 
(where $G_\Lambda \ne 0$)
is, in the limit $q \to 0$ with $\hat q = \vec{q}/|\vec{q}|$ fixed,
equal to 
$L_\Lambda^{(-)}=\{\vec{p} \in L_\Lambda: \vec{p}\cdot\hat{q} <0\}$,
which is the half of the Fermi surface.

Thus for the one--loop integrals, setting $\vec{q} = 0$ gives the same
result as the average over $\hat{q}$ of the limits $\vec{q} \to 0$ with $\hat q $
fixed. 
The above argument also shows
that seeming ambiguities in the limit are resolved by properly doing the
$H$--integrals. 
Looking at the details of the $H$ integration is also necessary for $q \ne 0$ 
if $q$ happens to be a perfect nesting vector for the Fermi surface.

If there is a van Hove singularity on the level set $L_\Lambda$, the
Jacobian $J(\Lambda,\theta)$ has a nonintegrable singularity
at each point $\vec{p}^*$ on $L_\Lambda$ where $\nabla \epsilon_{\vec{p}^*} = 0$. 
We cut out small neighborhoods of all such singularities by a smooth partition 
of unity. The above argument applies to the contributions from outside these
neighborhoods. In a small neighborhood of each saddle point, we can now change
variables from $\vec{p}$ to $x,y$ so that $\epsilon_{\vec{p}}-\Lambda = xy$. 
The Jacobian $j$ of this change of variables is smooth if the second derivative 
of $\epsilon$ at $\vec{p}^*$ is nondegenerate. The $X_+$ defined above now changes to 
\begin{equation}
\tilde X_+ =
\int_{x^2+y^2\le r^2} dx\, dy \; j(x,y) \; 
\frac{\eta^{-1} H'(\eta^{-1} xy)}%
{(ip_0 - \Lambda -xy -\Sigma_\Lambda(p_0,p(x,y)) \, H(\eta^{-1} xy)^2}
\;
G_\Lambda \Phi
\end{equation} 
One can now scale in any way one wants; the resulting integral diverges logarithmically
in $\eta$. Thus at these $\Lambda$, the right hand side of the flow equation diverges,
leading to an infinite slope in the solution as a function of the scale $\Lambda$.

\end{document}